\documentclass[11pt,a4paper]{article}

\usepackage[T1]{fontenc} 
\usepackage{microtype} 

\usepackage[hmarginratio=1:1,top=32mm,columnsep=20pt]{geometry} 

\usepackage{amsmath}
\usepackage{amssymb}
\usepackage{mathrsfs}
\usepackage{graphicx}
\usepackage[percent]{overpic}
\usepackage{latexsym}

\usepackage{abstract}

\usepackage{titlesec} 
\titleformat{\section}[block]{\large\bfseries\centering}{\thesection}{1em}{} 
\titleformat{\subsection}[block]{\bfseries}{\thesubsection}{1em}{} 
\numberwithin{equation}{section}

\usepackage{fancyhdr} 
\title{\vspace{-10mm}\fontsize{22pt}{10pt}\selectfont\textbf{Polarised antibranes from Smarr relations}\vspace{3mm}}

\author{
{Diego Cohen-Maldonado${}^1$, Juan Diaz${}^2$ and Fridrik Freyr Gautason${}^2$}\\[2mm] 
\small ${}^1$Institute of Physics, University of Amsterdam, Science park\\
\small Postbus 94485, 1090 GL Amsterdam, The Netherlands\\[2mm]
\small and\\[2mm]
\small ${}^2$Instituut voor Theoretische Fysica, KU Leuven\\
\small Celestijnenlaan 200D, 3001 Leuven, Belgium\\[3mm]
\texttt{\small\href{mailto:d.b.cohenmaldonado@uva.nl}{d.b.cohenmaldonado@uva.nl},
\small\href{mailto:juan.diaz@fys.kuleuven.be}{juan.diaz},
\small\href{mailto:ffg@fys.kuleuven.be}{ffg@fys.kuleuven.be}} 
}
\date{}

\usepackage{amsthm}

\usepackage{enumerate}
\numberwithin{equation}{section}

\usepackage{color}
\definecolor{dark-gray}{gray}{0.20}
\definecolor{gray}{gray}{0.30}
\definecolor{light-gray}{gray}{0.80}
\definecolor{dark-red}{rgb}{0.7,0,0}
\definecolor{dark-green}{rgb}{0.1,0.4,0}
\definecolor{dark-blue}{rgb}{0.3,0.3,0.7}
\definecolor{light-blue}{rgb}{0.8,0.8,1}
\definecolor{crapred}{rgb}{0.75,0,0}

\usepackage[super,sort&compress]{natbib}
\bibpunct{[}{]}{,\!}{n}{}{}
\usepackage{hyperref}
\usepackage{setspace}
\hypersetup{
	colorlinks=true,
	linkcolor=dark-blue,
	citecolor=dark-red,
	urlcolor=dark-green,
	linktoc=page
}

\newcommand{\dd}{\mathrm{d}}

\newcommand{\e}{\mathrm{e}}
\newcommand{\w}{\wedge}

\newcommand{\be}{\begin{equation}}
\newcommand{\ee}{\end{equation}}
\newcommand{\bea}{\begin{eqnarray*}}
\newcommand{\eea}{\end{eqnarray*}}

\newcommand{\f}[2]{\frac{#1}{#2}}
\newcommand{\R}{\mathbf{R}}

\newcommand{\vol}{\text{vol}}

\renewcommand{\d}{\text{d}}
\newcommand{\C}{\mathbb{C}}
\renewcommand{\R}{\mathbb{R}}

\begin{document}

\maketitle
\thispagestyle{fancy}

\begin{abstract}
We study the backreaction of smeared and localised anti M2-branes placed at the tip of the CGLP background. To this end we derive a Smarr relation for backreacted  antibranes at zero and finite temperature. For extremal antibranes we show that if smeared they cannot have regular horizons, whereas localised M2-branes can potentially be regular when polarised into M5-branes, in agreement with the probe result of Klebanov and Pufu. We further discuss antibranes at finite temperature and argue that localised antibrane solutions with regular horizons are not excluded.
\end{abstract}

\renewcommand{\baselinestretch}{0.9}\normalsize
\tableofcontents
\renewcommand{\baselinestretch}{1.0}\normalsize

\newpage
\section{Introduction}
Breaking supersymmetry in a controlled manner remains one of the hard challenges in constructing string theory 
vacua. A reliable mechanism would be welcome for many applications in string theory.
One way to make progress is to combine two objects which are both individually supersymmetric, while their combination is not.
In practice, one changes a sign of a charge such that the BPS conditions of these two elements are incompatible. 
Consider for example an M2-brane in flat space whose 16 Killing spinors obey
\begin{equation}
\Gamma_{012} \epsilon = \pm \epsilon~,
\end{equation}
where the sign reflects the charge of the brane. Combining two M2-branes with opposite charge results in 
incompatible BPS conditions and supersymmetry is broken. Such a solution is however also unstable; the branes 
attract each other and will eventually annihilate. The same game can be played by 
combining M2-branes with nontrivial flux
\begin{equation}
\dd G_7 = Q_\text{M2}\delta(\text{M2}) - \f12 G_4 \w G_4~.
\end{equation}
One then also breaks supersymmetry in much the same way as for two oppositely charged M2-branes. Brane/flux set-ups have already proved useful  in string cosmology \cite{Kachru:2003aw,Kachru:2003sx}, the black hole 
microstate program \cite{Bena:2012zi,Bena:2011fc} and dynamical supersymmetry breaking in holographic field theories \cite{Maldacena:2001pb,Kachru:2002gs,Argurio:2006ny,Argurio:2007qk,DeWolfe:2008zy}. Stability of 
such a background is however more complicated to analyse since the decay generically occurs through the
Myers effect \cite{Myers:1999ps}, with the M2-branes polarising into M5-branes that then subsequently 
decay. The M2-branes are then effectively annihilated against flux quanta in a process called brane/flux annihilation 
\cite{Kachru:2002gs}.

A concrete set-up involves placing anti M2-branes in the resolved cone background of Cveti{\v c}, Gibbons, L{\"u} and Pope (CGLP) 
\cite{Cvetic:2000db} analogous
to a similar one found by Klebanov and Strassler in type IIB supergravity \cite{Klebanov:2000hb}. This set-up was analysed
using a probe brane in \cite{Klebanov:2010qs} where it was found that the M2 would polarise to a spherical M5-brane
which finds a metastable state close to the original M2-brane location. Later, various approaches used to study the 
backreaction of M2-branes on the geometry revealed a divergent energy density for $G_4$ which could not be attributed 
to the presence of M2-branes \cite{Bena:2010gs,Massai:2011vi,Cottrell:2013asa,Giecold:2013pza,Blaback:2013hqa}.
One reasonable interpretation of the singularity is that because the M2s want to polarise to an M5-brane they induce
the observed singularity in $G_4$. The singular flux pile-up could signal that the brane/flux annihilation process 
occurs classically rather than through quantum tunnelling \cite{Blaback:2012nf,Danielsson:2014yga,Gautason:2015ola}.
If not, we would expect to be able to hide the singularity or any polarised metastable state 
behind a horizon by heating up the branes \cite{Gubser:2000nd}. Reference \cite{Blaback:2013hqa} found that \emph{smeared} antibranes 
exhibit a singular horizon at any temperature. A toy model analysis of localised branes showed that the result of \cite{Blaback:2013hqa}
might be an artifact of the smearing \cite{Hartnett:2015oda}. In this paper we aim to determine what is 
required so that a extremal polarised state exists, but we also revisit the non-extremal scenario.

Before discussing our results let us point out that the story outlined above closely follows a similar one for 
anti D3-branes in the Klebanov-Strassler background \cite{Kachru:2002gs}. These also exhibit a singularity that could not 
be attributed to the presence of D3-branes \cite{Bena:2009xk,McGuirk:2009xx,Bena:2011hz,Bena:2011wh,Bena:2012bk,Blaback:2011pn}. 
Many attempts to interpret the singularity have 
yielded negative results \cite{Bena:2012vz,Bena:2012ek,Bena:2013hr,Bena:2014jaa,Blaback:2014tfa,Bena:2015kia,Gautason:2015ola}. 
Nevertheless, a recent discussion has showed some promise in resolving the puzzle 
that the singularity raises. 
First, some have suggested that the singularity does not pose particular problems when only \emph{one}
anti brane is placed in KS as opposed to many \cite{Michel:2014lva,Bergshoeff:2015jxa}. Then it was discovered recently that a possible loophole exists
in previous arguments that allows for a polarised state and a finite temperature version of it that resolves the singularity \cite{Hartnett:2015oda,Cohen-Maldonado:2015ssa,Cohen-Maldonado:2015lyb}. 

In this paper, we study the backreaction of anti M2-branes and polarised M5-branes in the background of \cite{Cvetic:2000db}, using a technique employed in \cite{Gautason:2013zw} and further developed in \cite{Blaback:2014tfa,Cohen-Maldonado:2015ssa}. After reviewing the essential ingredients of the CGLP background in Section \ref{secCGLP}, we derive in Section \ref{secsmarr} the Smarr relation for a system of M2 and M5-branes placed at the tip of that geometry. It relates the energy measured far away from the branes to the charge and surface gravity of the M2/M5 system 
\be\label{smarr1}
\boxed{~{\cal E} = \f76 \f{\kappa {\cal A}}{8\pi G_N} + \Phi_\text{M2}Q_\text{M2}+\Phi_\text{M5}Q_\text{M5}~,~}
\ee
where $\kappa$ and ${\cal A}$ denote the surface gravity and area respectively, $\Phi_\text{M2}$ and $Q_\text{M2}$ denote the potential and M2-charge of the system, while $\Phi_\text{M5}$ and $Q_\text{M5}$ denote the \emph{dipole} potential
and charge of the brane system. Note that since the M5-branes are contractible, as we explain later, their monopole charge vanishes. However, we find similar to \cite{Emparan:2001wn,Cohen-Maldonado:2015ssa} that the dipole charge of the M5s contribute with a non-vanishing term to the Smarr relation. A non-vanishing dipole contribution is only possible for horizons with a non-trivial topology.\footnote{One could wonder whether the backreacted solution has no horizon but is rather supported by a topological contribution as in \cite{Gibbons:2013tqa}. In this paper we will not explore this possibility as we assume a presence of a horizon.} In Section \ref{secSmeared} we warm up by discussing smeared antibranes, which we will show cannot be regular. Then, in Section \ref{secExtremal}, we extend the results of \cite{Bena:2010gs,Massai:2011vi,Blaback:2013hqa} to localised branes, showing that extremal anti M2-branes with trivial horizon topology cannot have a regular horizon. If the horizon topology is non-trivial on the other hand, then the Smarr relation does not constrain the horizon to be singular. This is most likely the metastable state found by Klebanov and Pufu \cite{Klebanov:2010qs} although a full solution remains to be found. Finally in Section \ref{secBlack} we consider non-extremal branes. There we argue that localised branes posses at least two possible phases, differing  in their horizon topology, which we briefly discuss. We conclude with Section \ref{secConclusion}.

\section{Anti M2-branes in CGLP}\label{secCGLP}
In this section we review the smooth background of \cite{Cvetic:2000db} which is a warped product of $\R^{1,2}$ and a Stenzel manifold. 
We start with the construction of  Stenzel spaces before turning to the full supergravity field configuration.

Let us consider Calabi-Yau hypersurfaces in $\C^{n+1}$ with a conical singularity at the origin:
\be\label{cone}
{\cal C}^n = \big\{ z\in \C^{n+1}~:~ z_i z_i = 0\big\}~.
\ee
For $n\ge 3$, the base spaces of the cones are Sasaki-Einstein manifolds of dimension $2n-1$ and can be identified by intersecting ${\cal C}^n$ with
the unit sphere in $\C^{n+1}$:
\be
{\cal B}^{2n-1} = \big\{ z\in {\cal C}^n~:~z_i\bar{z}_i = 1\big\}~.
\ee
For $n=3$ the base space is ${\cal B}^5 = T^{1,1}$ whereas for $n=4$ the base is
${\cal B}^7 = V_{5,2}$. A resolution of the conical singularity of \eqref{cone} can be achieved by modifying the defining equation 
by adding an inhomogeneous term to the right hand side 
\be\label{defcone}
{\cal C}_\epsilon^n= \big\{ z\in \C^{n+1}~:~z_iz_i = \epsilon^2\big\}~,
\ee
where $\epsilon\in \R$.
For $n=2$ the hypersurface is the Eguchi-Hanson resolution of the $A_1$ singularity \cite{Eguchi:1978xp} while $n=3$ gives the well-known 
deformed conifold \cite{Candelas:1989js}. The explicit metrics can be derived using a K{\"a}hler potential $K$ which only depends on the variable
\begin{equation}
\rho = z_i\bar{z}_i~,
\end{equation}
and  satisfies the differential equation \cite{stenzel}
\begin{equation}
\rho (K')^n + (\rho^2-\epsilon^4)K'' (K')^{n-1}=c^2~,
\end{equation}
for some normalization constant $c$. After solving this equation the metric can be written down
\be\label{stenzel}
\dd s_{2n}^2 = K'(\rho) \sum_{i=1}^{n+1}\dd z_i\dd \bar{z}_i + K''(\rho)\sum_{i=1}^{n+1} |z_i \dd \bar{z}_i|^2~.
\ee

We will focus exclusively on $n=4$ with $c = 9/4\epsilon^3$ for which an explicit form of the metric can be found in \cite{Klebanov:2010qs}. 
Since we do not 
require its explicit form in this paper we omit writing it. From now on we will rescale our coordinates to absorb $\epsilon$, then
 the coordinate $\rho$ ranges between $1$ and 
$\infty$ and for large $\rho$ the metric reduces to that of the cone \eqref{cone}. Finally, for $\rho=1$ the metric reduces to that of a 4-sphere
with radius $\sqrt{3/2}$.

\subsection{The CGLP background}\label{SecAction}
The supergravity background of \cite{Cvetic:2000db} is a warped product of the metric \eqref{stenzel} with $n=4$ and flat 3-dimensional 
spacetime:
\be\label{CGLPmetric}
\dd s^2 = H^{-2/3}\big(-\dd t^2 + (\dd x^1)^2 + (\dd x^2)^2\big) + H^{1/3}\dd s_8^2~.
\ee
This metric solves the Einstein equation derived from the action of 11-dimensional supergravity
\begin{equation}\label{11daction}
S=\f{1}{16\pi G_N}\int \left\{\star_{11} R-\frac{1}{2}\star_{11}G_4\wedge G_4-\frac{1}{6}G_4\wedge G_4\wedge A_3\right\}~,
\end{equation}
where the form fields are
\begin{align}\label{CGLPfs}
G_4 &= -\vol_3\w\dd H^{-1} + m \omega_4\\
G_7 &\equiv \star_{11}G_4 = H^2 \star_8\d H^{-1}-m H^{-1}\vol_3\wedge \omega_4~.
\end{align}
Here $\vol_3 = \dd t\w\dd x^1\w\dd x^2$ and $\omega_4$ is an anti self-dual closed 4-form on ${\cal C}_\epsilon^4$, $m$ is a constant and $\star_8$ is the Hodge operator on $\dd s_8^2$. The Bianchi identity $\dd G_4 =0$ is 
trivially solved for \eqref{CGLPfs} whereas the Bianchi identity for $G_7$:
\begin{equation}
\dd G_7+\frac{1}{2}G_4\wedge G_4=0~,
\end{equation}
implies
\be
\dd \star_8 \dd H =\f12 m^2 \star_8\omega_4\w\omega_4~,
\ee
which can be written as
\be
\nabla_8^2 H =-\f12 m^2 |\omega_4|^2~.
\ee
This equation can be integrated assuming $H\equiv H(\rho)$ and regularity at $\rho=1$ \cite{Cvetic:2000db,Klebanov:2010qs}
\be \label{background-warp}
H = c_0 + \big(12^3 ~3^{8}\big)^{1/4} m^2\int_{(2+\rho)^{1/4}}^{\infty}\f{\dd t}{(t^4-1)^{5/2}}~.
\ee
The constant $c_0$ controls the asymptotic behaviour of the solution. We will consider both $c_0=0$ for which the metric is asymptotically
AdS$_4\times V_{5,2}$, and $c_0\ne 0$ for which the solution is asymptotically Ricci flat, $\R^{1,2}\times {\cal C}^4$. For $c_0\ne 0$ we can rescale
the coordinates as well as $m$ to absorb $c_0$. Therefore we will only consider $c_0=1$ in addition to $c_0=0$.

\subsection{Probe anti M2-branes}
Anti M2-branes placed in the M-theory background just described experience a radial force which pulls them towards the resolved tip of the cone. In \cite{Klebanov:2010qs} Klebanov and Pufu performed a probe analysis to determine the behaviour of $p$ anti M2-branes sitting at the tip. In this section we review their results.

Locally, close to the tip, the metric \eqref{stenzel} reduces to the metric on the 4-sphere
\be
\d s^2_8 \underset{\rho\to1}{\longrightarrow}  \f{3}{2}\left[\dd\psi^2 + \sin^2\psi \dd \Omega_3^2\right]~,
\ee
where $\dd \Omega^2_3$ is the metric on the round three-sphere and $\psi\in[0,\pi]$ is the azimuthal angle on the four-sphere. Without loss of generality, one may assume that the antibranes are initially located at the North pole, with $\psi=0$. The interaction between the branes and the background flux gives rise to a polarisation process through the Myers effect. Concretely, the anti M2-branes polarise into an M5-brane carrying finite M2 charge wrapping a finite size $S^3$ at a certain value of $\psi$.

The probe calculation follows closely the initial work of \cite{Kachru:2002gs}. By evaluating the Lagrangian of a probe M5-brane with $p$ units of M2 charge, one obtains an effective potential as a function of the azimuthal angle and the ratio $p/M$, where $M$ is the total $G_4$ flux threading the four-sphere:\footnote{We use conventions where $16 \pi G_N = (2\pi)^8 \ell_p^9$.}
\begin{equation}\label{4-cycle charge}
M = \frac{1}{(2\pi \ell_p)^3}\int_{S^4}G_4=\frac{18 \pi^2 m}{(2\pi \ell_p)^3}~.
\end{equation}
Depending on the value of $p/M$, this potential has either only one absolute minimum at $\psi=\pi$, corresponding to the supersymmetric state where the M5-brane has $p-M$ units of M2-brane charge which preserves the same supersymmetry as the flux background, or one absolute minimum at $\psi=\pi$ plus a local minimum at some value $\psi=\psi_{\text{min}}$, corresponding to a metastable polarised state.

\begin{figure}[h!]
\begin{center}
\includegraphics[width=0.8\textwidth]{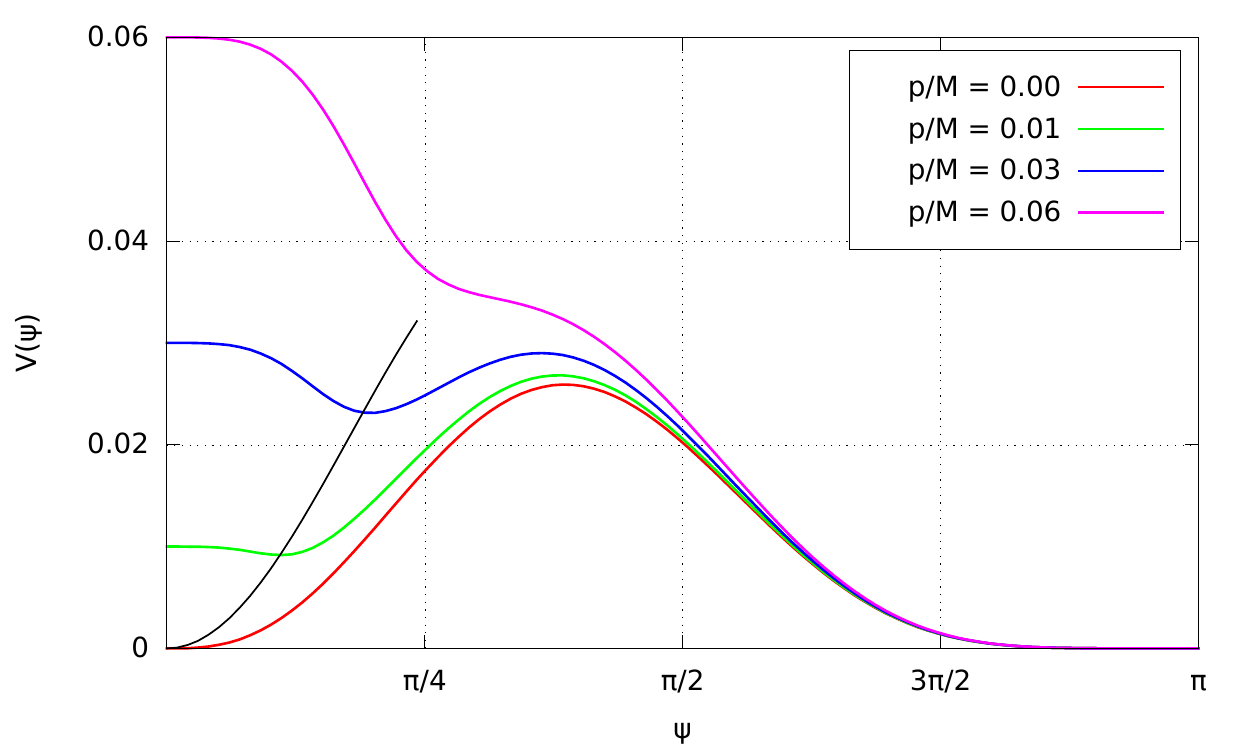}
\caption{The polarisation potential $V(\psi)$ for different values of $p/M$. The black line shows the estimate for the position of the metastable minimum given in eq. \eqref{Pol-angle}. \label{fig-potential}}
\end{center}
\end{figure}

The analysis of \cite{Klebanov:2010qs} was carried out after a dimensional reduction along one of the coordinates of the anti M2-branes. We then have anti fundamental strings in type IIA supergravity, which polarise into D4-branes. The polarisation angle $\psi_{\text{min}}$ is found at the minimum of the polarisation potential
\begin{equation}
V(\psi)=\sqrt{h \sin^6\psi+U^2(\psi)} - U(\psi)~,
\end{equation}
where
\begin{equation}
U(\psi)=\frac{1}{8}\cos^3\psi - \f38 \cos\psi + \frac{1}{4} - \f{p}{2M}~,
\end{equation}
and
\be
h = \f{H(1)}{96m^2} = \f{c_0}{96m^2} + \left(\f{3}{4}\right)^{7/4}\int_{3^{1/4}}^{\infty}\f{\dd t}{(t^4-1)^{5/2}}
\approx 0.0114~,
\ee
where we used that $m\gg 1$.
In Figure \ref{fig-potential} we plot the polarisation potential for different values of $p/M$. A metastable minimum of the potential only exists for small range of parameters $0< p/M\lesssim 0.0538$.
Furthermore, for small $p/M$  the minimum satisfies
\begin{equation}\label{Pol-angle}
\psi_{\text{min}}^2 \approx \frac{1}{8 h}\frac{p}{M}~.
\end{equation}

\section{A Smarr relation for M2-branes}\label{secsmarr}
In this section we derive the Smarr relation \eqref{smarr1} for $p$ anti M2-branes in the CGLP background with flux number $M$. We will assume that $p/M\ll 1$, in line with \cite{Klebanov:2010qs}. In Sections \ref{secSmeared}, \ref{secExtremal} and \ref{secBlack} we then use this formula to constrain both extremal and non-extremal antibrane solutions. We find that smeared and extremal pointlike anti M2-branes are not consistent with the Smarr relation, whereas polarised and non-extremal states are. Our approach is reminiscent of the one employed in \cite{Gautason:2013zw} and later \cite{Blaback:2014tfa} for type II antibrane systems. In Appendix \ref{appendixDbranes} we derive the Smarr relation for such set-ups.

In order to perform the calculation, we make an important assumption that asymptotically, far away from the tip of the cone, the solution should look like the CGLP solution described above. In particular the M2 charge measured at infinity is fixed to the one for a CGLP background for a given $m$.  When the antibranes are introduced we adjust $m$ so that the charge remains the same. This will lead to a nonvanishing ADM mass measured at infinity as we will explain in Section \ref{secExtremal}. It is the aim of this section to obtain the Smarr relation between the ADM energy density, area, charge and chemical potentials of the antibrane system.

The full antibrane metric is assumed to take the form
\begin{equation}\label{m2metric}
\dd s^2_{11}=e^{2A}\left(-\e^{2f}\dd t^2 + (\dd x^1)^2+(\dd x^2)^2\right)+ \d \bar{s}_8^2~,
\end{equation}
where $\d \bar{s}^2_8$ is a modification of the metric on ${\cal C}_\epsilon^4$ which takes into account the backreaction of the M2 branes on the flux background. We omit writing an explicit warp factor in front of the 8-dimensional metric but assume that asymptotically, for large $\rho$,
\begin{equation}
\dd \bar{s}_8 \to H^{1/3}\dd s_8^2~,
\end{equation}
where $\dd s_8^2$ is given in eq. \eqref{stenzel}.
Note that we have introduced a metric function $\e^{2f}$ that breaks the Lorentz symmetry of $\R^{1,2}$ to incorporate a possible non-extremal state.
The metric is general enough to describe either a stack of anti M2-branes or polarised M5-branes carrying M2 charge. 

In the following it will be useful to introduce notation for the gauge fields that is adapted to the metric \eqref{m2metric}.
We write
\begin{align}
G_4&=-\e^{3A+f}\vol_3\wedge F_1+\tilde{F}_4~, \label{g4def}\\
G_7&=\e^{3A+f}\vol_3\wedge F_4+ F_7~,\label{g7def}
\end{align}
which implies $F_4 = \star_8 \tilde{F}_4$ and $F_7 = -\star_8 F_1$. In this section $\star_8$ refers to the Hodge operator on $\dd\bar{s}_8^2$.
With these definitions the equations of motion in absence of source terms take the form
\begin{eqnarray}
\d F_7+\frac{1}{2}\tilde{F}_4\wedge \tilde{F}_4 &=&0~,\label{formeq1}\\
\d(e^{3A+f}F_4)+e^{3A+f}F_1\wedge\tilde{F}_4&=&0~,\label{formeq2}\\
\d(e^{3A+f}F_1)&=&0~,\label{formeq3}\\
\d\tilde{F}_4&=&0~.\label{formeq4}
\end{eqnarray}
For pointlike M2-branes in the internal space, only eq. \eqref{formeq1} receives a delta function contribution on the right hand side. For M5-branes that
wrap three internal dimensions only eq. \eqref{formeq4} receives a contribution, unless the M5 carries M2 charge which then disguises itself as a contribution to eq. \eqref{formeq1}.
For large $\rho$, the asymptotic expansion of
all field strengths and warp factor should equal the one for the CGLP background to leading order. Beyond leading order, fields will generically differ from their background values.
We therefore let
\be\label{asymptotics1}
\e^{2f}\to 1~,\quad \e^{2A}\to H^{-2/3}~,
\ee
as $\rho\to\infty$.

We will from now on assume that there are globally well-defined gauge potentials for $F_4$ and $F_1$, defined by
\begin{equation}\label{potentials}
e^{3A+f}F_1=\d C_0 \quad \quad \text{and} \quad \quad e^{3A+f}F_4=\d C_3-C_0\tilde{F}_4~.
\end{equation}
Despite the suggestive notation, $C_0$ appears in the Wess-Zumino (WZ) terms for M2-branes while $C_3$ appears in the WZ terms for M5-branes.
In the limit $\rho\to\infty$ the potentials reduce to 
\be\label{asymptotics2}
C_0\to H^{-1}\quad\text{and}\quad C_3\to 0~,
\ee
and so these are globally defined for the CGLP background. The presence of M2-branes or their polarised state does not affect the existence of $C_0$ and $C_3$ in line with the discussion below eqs. (\ref{formeq1}-\ref{formeq4}).
The gauge transformations that leave the field strengths invariant are
\begin{equation}
\delta C_0 =0~,\quad \delta C_3 = \dd \lambda_2~.
\end{equation}

\subsection{ADM energy}
We now turn our attention to the ADM energy density of the anti M2-branes. We will relate it to the potentials $C_0$ and $C_3$ evaluated at the 
horizon of the brane configuration. The general expression for the ADM energy density of a $p$-brane configuration in $D$ dimensions
is derived in Appendix \ref{appendixadm}, which extends the results of \cite{Townsend:2001rg} to spacetimes which are not transverse 
asymptotically Ricci flat. The result is
\be\label{genEnergy}
{\cal E} = -\f{1}{16\pi G_N} \oint_\infty \star_D \left[\dd \eta\w\Lambda_p + \xi\w\eta\w\Lambda_p 
+ \f{1}{D-p-3}\dd(\eta\w\Lambda_{p})\right]~,
\ee
where $\Lambda_p=\lambda_{(1)}\w\cdots\w\lambda_{(p)}$, $\eta$ is a one-form dual to the timelike Killing vector $\partial_t$ and $\lambda_{(i)}$ are one-forms dual to $p$ spacelike killing vectors $\partial_{x^i}$, $i=1,\ldots,p$. Finally, $\xi$ is a one-form that takes care of subtracting the background contribution to the energy density and corresponds to adding a counter-term to the action
\be
\int \dd \star_D \xi~.
\ee
We normalize the energy with respect to the CGLP background for which (see Appendix \ref{appendixadm})
\be
\xi = \dd \log H~.
\ee
Using this \eqref{genEnergy} reduces to 
\be\label{adm1}
{\cal E} = \f{1}{16\pi G_N}\f13 \oint_\infty \e^{3A+f} \star_8\dd(9A+7f+3\log H)~.
\ee
We now use the Einstein equation to bring the integration surface from $\rho\to\infty$ close to the horizon of the branes. To this end we
write the components of the Einstein equation along the brane worldvolume,
\be \label{einstein}
R_{\mu\nu} + \f16 g_{\mu\nu}\left(2|F_7|^2 + |\tilde F_4|^2\right) = 0~,
\ee
for $\mu,\nu=0,1,2$. Using the form of the metric \eqref{m2metric} the Ricci tensor on $\R^{1,2}$ can be explicitly written down,
\begin{align}
R_{00} &= - g_{00}\left[\square_8(A+f) + \nabla(3A+f)\cdot \nabla(A+f)\right]~,\\
R_{ij} &= - g_{ij}\left[\square_8 A + \nabla(3A+f)\cdot \nabla A\right]~,
\end{align}
where $i,j=1,2$ and the dot-product is performed using the transverse metric $\dd \bar{s}_8^2$. The Einstein equation \eqref{einstein} reduces to two differential equations that will enable us to rewrite the ADM energy,
\begin{eqnarray}
\dd\left(\e^{3A+f}\star_8 \dd f\right)&=&0~,\\
\dd\left(\e^{3A+f}\star_8 \dd A\right) &=& -\f{\e^{3A+f}}{3}\star_8\left(|F_7|^2 + \f12 |\tilde F_4|^2\right)~.
\end{eqnarray}
We define an 8-dimensional submanifold ${\cal M}_8$ that has boundaries at $\rho\to\infty$ and at the horizon of the brane configuration. Using the
above differential equations together with \eqref{adm1} yields
\begin{eqnarray}
{\cal E} &=& \f{1}{16\pi G_N}\f13 \oint_H \e^{3A+f} \star_8\dd(9A+7f)\nonumber\\
&& - \f{1}{16\pi G_N}\int_{{\cal M}_8}\e^{3A+f}\star_8\left(|F_7|^2 + \f12 |\tilde F_4|^2\right)\label{adm2}\\
&&+\f{1}{16\pi G_N}\oint_\infty \e^{3A+f}\star_8 \dd\log H~,\nonumber
\end{eqnarray}
where the subscript $H$ in the first term denotes the horizon.

We will analyse the three terms of \eqref{adm2} individually. First, by construction the warp factor $A$ is completely regular at the horizon and\footnote{Note that for extremal horizons $A$ diverges, wheras $f$ vanishes. It is simple to verify that all results obtained in this section are valid also for extremal horizons taking the limit $f\to 0$.}
\begin{equation}
\e^{3A+f}\to 0\quad \text{as}\quad \rho\to \rho_H~.
\end{equation}
This implies that we can rewrite the first term of \eqref{adm2} as
\be
\f13 \oint_H \e^{3A+f} \star_8\dd(9A+7f) 
= -\f76\oint_H \star_{11}\dd \eta\w\lambda_{(1)}\w\lambda_{(2)}~.
\ee
The integral on the right-hand side above yields exactly minus two times the horizon surface gravity $\kappa$ times 
the \textit{effective area} of the horizon ${\cal A}$ \cite{Myers:1986un,Townsend:2001rg} (see also Appendix \ref{appendixArea}), and so 
\be\label{TSterm}
\f{1}{16\pi G_N}\f13 \oint_H \e^{3A+f} \star_8\dd(9A+7f) = \f{7}{6}\f{\kappa {\cal A}}{8\pi G_N}~.
\ee
The second term of eq. \eqref{adm2} can be rewritten using (\ref{formeq1}-\ref{formeq3}) together with the definitions \eqref{potentials}
\be\label{rewritingeinstein}
\e^{3A+f}\star_8\left(2|F_7|^2 + |\tilde F_4|^2\right) = \dd(C_3\w \tilde{F}_4 + 2C_0F_7)~.
\ee
The bulk integral can therefore be transformed into a surface integral evaluated at the horizon and at infinity. 
At infinity we can use the behaviour of the fields (\ref{asymptotics1}-\ref{asymptotics2}) so that
\be\label{tobecountered}
-\f{1}{16\pi G_N}\oint_\infty \left(\f12 C_3\w \tilde{F}_4 + C_0F_7\right) = -\f{1}{16\pi G_N}\oint_\infty  C_0 F_7=-Q_\text{M2}~\lim_{\rho\to\infty}H^{-1}(\rho) ~, 
\ee
where\footnote{In our units the charge of a single M2-brane is $\mu_\text{M2}= 2\pi/(2\pi \ell_p)^3$ and the unit charge of an M5-brane is
$\mu_\text{M5} = 2\pi/(2\pi \ell_p)^6$.}
\begin{equation}
Q_\text{CGLP} = \f{\mu_\text{M2}}{(2\pi\ell_p)^6}\int F_7 = \mu_\text{M2}\f{M^2}{4}~,
\end{equation}
is the M2 charge of our solution which we assume to be the same as the one of the CGLP background (see Appendix \ref{CGLPcharge}). When $c_0=0$
the term \eqref{tobecountered} diverges as $\rho^{9/4}$ but is exactly cancelled by the last term in \eqref{adm2}
\be
\f{1}{16\pi G_N}\oint_\infty \e^{3A+f}\star_8 \dd \log H = Q_\text{CGLP}~\lim_{\rho\to\infty}H^{-1}(\rho)~.
\ee

Combining the above results we can write the ADM energy as
\be\label{ADM-energy}
{\cal E} = \f{7}{6} \f{\kappa {\cal A}}{8\pi G_N} + \f{1}{16\pi G_N}\oint_H \left(\f12 C_3\w \tilde{F}_4 + C_0F_7\right)~.
\ee
For the CGLP background the horizon area vanishes and the regularity of the background ensures that the second integral also vanishes so that we end
up with the expected result
\begin{equation}
{\cal E}_\text{CGLP} = 0~.
\end{equation}
The equation \eqref{ADM-energy} has non-trivial implications for the consistency of the supergravity solutions describing anti M2-branes and polarised M5-branes at the tip of the cone. It allows us to relate the UV behaviour of the solution, characterized by the ADM energy measured at infinity, to the IR structure of the horizon. In \cite{Blaback:2014tfa,Cohen-Maldonado:2015ssa} such a  relation was used to argue for a singular flux at the horizon of localised anti D3-branes sitting at the tip of the Klebanov-Strassler background as a result of demanding a non-vanishing ADM energy. However, we will use \eqref{ADM-energy} in a somewhat different way: we will assume that the solutions have regular horizons, and from there on investigate what it implies for the ADM energy measured in the UV.

\subsection{Charges and potentials}
We now close the ADM discussion by interpreting the last term in \eqref{ADM-energy}. First of all, from the equations of motion \eqref{formeq1}-\eqref{formeq4} we can write a local gauge potential for $\tilde{F}_4$:
\begin{equation}
\tilde{F}_4=\d \mathcal{H}_3~, \quad \quad F_7=\tilde{F}_7-\frac{1}{2}\mathcal{H}_3\wedge \tilde{F}_4~,
\end{equation}
where $\tilde{F}_7$ is a closed 7-form. With this we can rewrite the horizon integral as
\begin{equation}\label{horizonInt}
\oint_H \left(\f12 (C_3-C_0\mathcal{H}_3)\w \tilde{F}_4 + C_0\tilde{F}_7\right)~.
\end{equation}
The integral involving $C_0$ and $\tilde{F}_7$ has the structure of the potential-charge ($\Phi_\text{M2} Q_\text{M2}$) term that is standard in Smarr relations for black holes. Indeed, we will see that $C_0$ is constant at the horizon. Moreover, the integral of $\tilde{F}_7$ corresponds to the Page charge sourced by the branes, and hence measures the localised M2 charge present in the geometry through
\begin{equation}
Q_{\text{M2}}=\frac{\mu_{\text{M2}}}{(2\pi \ell_p)^6}\oint\tilde{F}_7.
\end{equation}
We are left with
\begin{equation}
 \f{1}{16\pi G_N}\oint_H C_0\tilde{F}_7=\Phi_{\text{M2}}Q_{\text{M2}},
\end{equation}
where $\Phi_{\text{M2}}$ equals to the gauge potential $C_0$ evaluated at the horizon:
\begin{equation}
\Phi_{\text{M2}}=C_0|_H~.
\end{equation}

As for the other term in the integral \eqref{horizonInt}, we will now argue that the three form $(C_3-C_0\mathcal{H}_3)$ restricted to the horizon is closed. The Einstein equation for 11-dimensional supergravity takes the form
\begin{equation}
R_{\mu\nu}-\frac{1}{2\cdot 3!}G_{\mu\rho_1\rho_2\rho_3}G_\nu^{\phantom{\nu}\rho_1\rho_2\rho_3}+\frac{1}{6}g_{\mu\nu}|G_4|^2=0~,
\end{equation}
from which we derived eq. \eqref{einstein}. At the Killing horizon of the timelike Killing vector $\xi$ we have\footnote{The second equality follows from the Raychaudhuri equation.}
\begin{equation}
|\xi|^2 = 0~\quad\text{and}\quad \xi^\mu\xi^\nu R_{\mu\nu} = 0~.
\end{equation}
Contracting the Einstein equation with $\xi$ at the horizon yields
\begin{equation}\label{juaneq}
|\iota_\xi G_4|^2 = 0~.
\end{equation}
Analogously we can write the Einstein equation in terms of the dual field strength $G_7$ and run the same argument to show that at the horizon
\be
|\iota_\xi G_7|^2 = 0~.
\ee
Using the definitions \eqref{g4def} and \eqref{g7def}, we can rewrite the previous equations as:
\begin{align}
\e^{-4A}|\e^{3A+f}F_1|^2 &= \e^{-4A}|\d C_0|^2= 0~, \\
\e^{-4A}|\e^{3A+f}F_4|^2& = \e^{-4A}|\dd C_3 - C_0 \tilde F_4|^2= 0~.\label{normf4vanishes}
\end{align}
It follows that $C_0$ is constant along the horizon as stated before and furthermore that $C_3-C_0\mathcal{H}_3$ restricted to the horizon is closed. The latter allows us to write
\be
C_3-C_0\mathcal{H}_3 = \omega_3 + \text{exact}~,
\ee
where $\omega_3$ is harmonic at the horizon. Furthermore since $\dd \tilde F_4=0$, the integral of $(C_3-C_0\mathcal{H}_3)\w \tilde F_4$ reduces to
\be\label{derivedipole}
\oint_H (C_3-C_0\mathcal{H}_3)\w \tilde F_4 = \oint_H \omega_3 \w \tilde F_4 = \f{(2\pi\ell_p)^9}{\pi}~\Phi_\text{M5}Q_\text{M5}~.
\ee
Here the M5-charge $Q_\text{M5}$ is defined by
\be\label{Q5def}
Q_\text{M5} = \f{\mu_\text{M5}}{(2\pi\ell_p)^3} \int_{{\cal M}_4}\tilde F_4~,
\ee
where ${\cal M}_4$ is a submanifold of the horizon which is related to the  Poincar{\'e} dual of $\omega_3$ by a constant of proportionality $2\Phi_\text{M5}$. We are now in position to write Smarr's relation for a system of anti M2-branes normalised for the background energy of the CGLP background:
\be\label{smarr}
{\cal E} = \f76 \f{\kappa {\cal A}}{8\pi G_N} + \Phi_\text{M2}Q_\text{M2} + \Phi_\text{M5}Q_\text{M5}~.
\ee
The numerical factor $7/6$ seems rather ad-hoc in this equation but is correct. We can see this by deriving the first law of black hole thermodynamics. It is easy to verify that $\kappa$ scales with the area in a non-trivial way
\begin{equation}
[\kappa] = L^{-1} = [{\cal A}]^{-1/7}~,
\end{equation}
whereas the chemical potentials do not scale with the charge. Using this, the first law takes the expected form
\begin{equation}
\dd {\cal E} =  \f{\kappa}{8\pi G_N}\dd{\cal A} + \Phi_\text{M2}~\dd Q_\text{M2} + \Phi_\text{M5}~\dd Q_\text{M5}~.
\end{equation}

\subsection{Relation to on-shell brane actions}
In \cite{Gautason:2013zw} a similar relation between brane charges and the cosmological constant of a compactification of type II supergravity was obtained. There the derivation relied on using delta functions in the equations of motion, which result from varying the brane worldvolume action. This is only relevant for extremal branes for which the worldvolume actions are known. We can also do this in 11-dimensional supergravity where the modified form equations of motion take the form
\begin{eqnarray}
\dd F_7 + \f12 \tilde F_4\w \tilde F_4 &=& Q_\text{M2}\delta_8 - Q_\text{M5} {\cal F}_3 \w \delta_5~,\\
\dd \tilde F_4 &=& -Q_\text{M5} \delta_5~.
\end{eqnarray}
In these equations ${\cal F}_3$ is the self-dual tensor field living on the M5 brane. It is fixed by gauge invariance of the M5 action to be ${\cal F}_3 = \dd b_2 + A_3$ with $b_2$ a 2-form and $A_3$ the gauge potential for $G_4$. The Einstein equation will also receive delta function contributions from the DBI actions of the branes but since we only use its external components in the derivation of the ADM energy, we only need to consider the couplings to form fields. 

We can now repeat the evaluation of the ADM energy using delta functions in the equations of motion, following closely the calculation performed in the last two subsections.
All equations remain unchanged up to \eqref{rewritingeinstein}, which now takes the form:
\begin{eqnarray}
\e^{3A+f}\star_8\left(2|F_7|^2 + |\tilde F_4|^2\right) &=& \dd(C_3\w \tilde{F}_4 + 2C_0F_7)\nonumber\\
&& - (C_3-2{\cal F}_3 C_0)\w Q_\text{M5}\delta_5 - 2C_0 Q_\text{M2} \delta_8~,\label{formeqwithdelta}
\end{eqnarray}
where $b_2$ is assumed to only have legs transverse to the M2 worldvolume.
The first term of equation \eqref{adm2} is zero since we only discuss extremal branes, the second one reduces to an integral 
over the delta functions after cancelling the infinite contribution using the third term:
\be
{\cal E} =  Q_\text{M2}\int C_0  \delta_8 + \f{Q_\text{M5}}{2}\int (C_3-2{\cal F}_3 C_0)\w \delta_5~.
\ee
Note that we have assumed that the delta functions take care of the singularities and that the total derivative in \eqref{formeqwithdelta} is free of any singularities. Identifying the chemical potential $\Phi_\text{M2}$ with $C_0$ and $\Phi_\text{M5}$ with the integral
\be
\Phi_\text{M5}=\f12 \int (C_3-2{\cal F}_3 C_0)\w \delta_5~,
\ee
we reproduce the Smarr relation \eqref{smarr}. It is interesting to note that the Smarr relation has the form of a sum of on-shell brane actions in analogy with the results of \cite{Gautason:2013zw}. A recent paper has suggested that this is not an accident and in general the on-shell actions of branes will arise in the calculation of the on-shell gravitational action (or free energy) of a given system \cite{Ferrari:2016vcl}.

\section{Smeared anti M2-branes}\label{secSmeared}
As a warm-up we will start by considering smeared antibranes. Smeared branes preserve the full SO$(5)$ symmetry of the 4-sphere at the tip of the background. This implies that the gauge potential $C_3$ vanishes. Regularity of the horizon then implies that 
\be
\Phi_\text{M2} = C_0|_H= 0~,
\ee
as follows from eq. \eqref{normf4vanishes}. Finally, it easy to verify that a Smarr relation for smeared branes cannot have a dipole contribution. This follows from eq. \eqref{derivedipole} together with the previous result that $C_3=C_0=0$. The final Smarr relation for smeared branes then takes the form 
\be\label{smearedsmarr}
{\cal E} = \f76 \f{\kappa{\cal A}}{8\pi G_N}~.
\ee
Such a Smarr relation cannot be attributed to branes with antibrane charge. In particular an extremal limit would give zero energy to the Smarr relation which cannot represent a stack of supersymmetry breaking antibranes sitting at the tip of the geometry.

This result has previously been observed as singular backreaction of the antibranes on the flux background \cite{Bena:2010gs,Massai:2011vi,Blaback:2013hqa}. Our calculation does not allow for such a singularity since we assumed a regular horizon. If we would not have done so, then we could not conclude that $\Phi_\text{M2}$ vanishes, but we would then also see that the solution exhibits the previously found singularity.

\section{Extremal anti M2-branes}\label{secExtremal}
In this section we use eq. \eqref{smarr} to constrain localised extremal antibrane solutions. Extremality implies that \eqref{smarr} reduces to
\be\label{smarrextr}
{\cal E} = \Phi_\text{M2}Q_\text{M2} + \Phi_\text{M5}Q_\text{M5}~.
\ee
All quantities on the right hand side of \eqref{smarrextr} are evaluated in the limit of zero horizon area.
For the set-up in question, the ADM energy measured at the UV is proportional to two times the red-shifted tension of the $p$ anti M2-branes sitting at the tip of the throat \cite{Klebanov:2010qs},
\begin{equation}\label{twotension}
\mathcal{E} = 2pT_{\text{M2}}e^{3A/2}~.
\end{equation}
Here $\e^{3A}$ is the red-shift factor generated by the warping of the background evaluated at the tip. This equation can be understood as follows. We fix the M2-charge at infinity to be the same as for a CGLP background with a given $m$. This charge is calculated in Appendix \ref{CGLPcharge} and appears in equation \eqref{M2charge}. For every anti M2-brane introduced into the background, $m$ must be adjusted so that the charge remains constant. This is equivalent to adding an M2-brane together with every anti M2-brane at the tip of the geometry which explains the factor of 2 in eq. \eqref{twotension}.

There is the further constraint from eq. \eqref{normf4vanishes} that restricted to the horizon, 
\be\label{extremalcancellation}
\dd C_3 = C_0 \tilde F_4~.
\ee
We now focus on the component of this equation along the 4-sphere at the tip. $\tilde{F}_4$ must be proportional to the volume form on the 4-sphere at the tip, since its integral there is proportional to $M$. The symmetries of the solution require that the only component of $C_3$ along the 4-sphere takes the form 
\be
f(\rho,\psi)\vol_{S^3}
\ee
for a function $f$ of the cone coordinate $\rho$ and the azimuthal angle $\psi$ on the 4-sphere with the antibranes sitting at $\psi =0$. Since $C_3$ is globally defined by construction, we conclude that $f(\rho,\psi)$ should reach either a minimum or a maximum at the poles, and therefore $\d C_3$ restricted to $\psi=0$ at the tip vanishes. Then eq. \eqref{extremalcancellation} yields:
\begin{equation}
C_0|_H= \Phi_\text{M2}=0
\end{equation}
for pointlike antibranes. The conclusion is that the first term in the right-hand side of \eqref{smarrextr} cannot contribute and the Smarr relation reduces to 
\be
\mathcal{E} = \Phi_\text{M5}Q_\text{M5}~.
\ee
Moreover, it is simple to see that for a pointlike horizon, just like for a smeared one, the M5-charge $Q_\text{M5}$ as defined in \eqref{Q5def} is zero. This can be seen by freely transforming the integration domain in the definition of $Q_\text{M5}$ in \eqref{Q5def} to infinity using the fact that $\tilde F_4$ is closed. Since we demand CGLP asymptotics, and therefore no M5 charge as measured at infinity, we obtain
\begin{equation}
Q_\text{M5}=0~.
\end{equation}
We conclude that there is no way to satisfy the Smarr relation \eqref{smarrextr} for pointlike anti M2-branes present at the tip.

\begin{figure}[h!]
\begin{center}
\begin{overpic}[width=0.3\textwidth,tics=10]{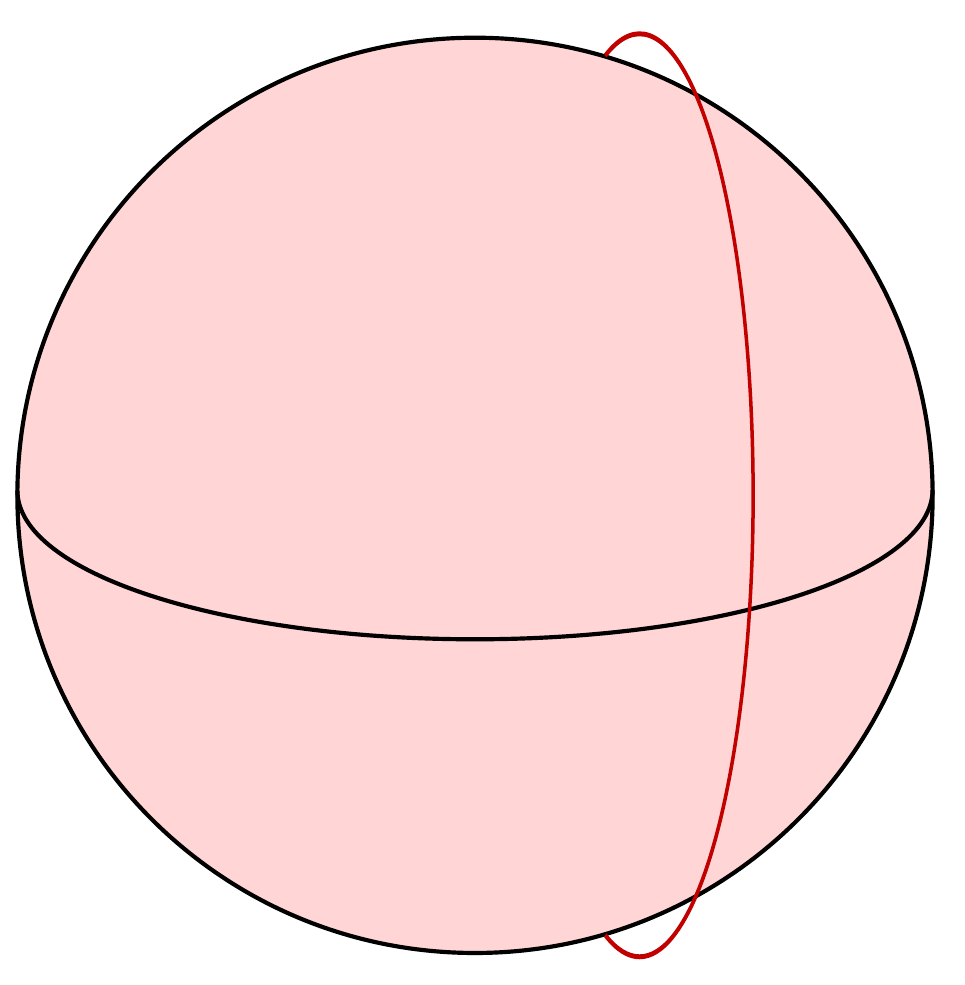}
\put(60,47){\textcolor{crapred}{${\cal M}_4$}}
\end{overpic}
\hspace{1.2cm}
\begin{overpic}[width=0.5\textwidth,tics=10]{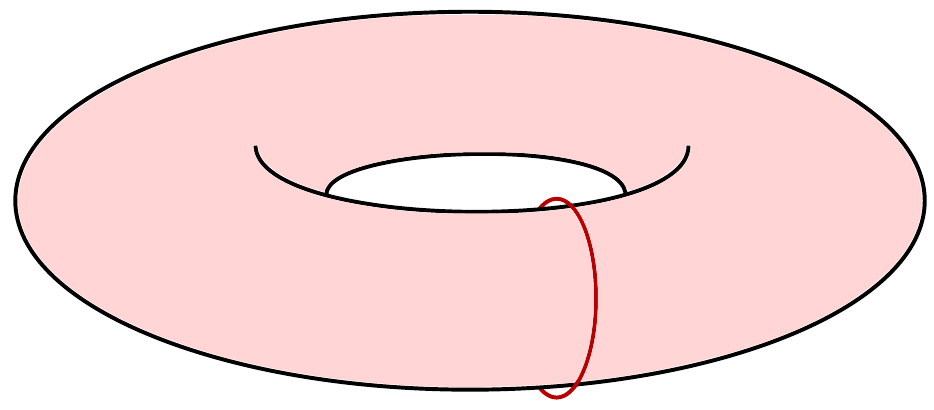}
\put(52,10){\textcolor{crapred}{${\cal M}_4$}}
\end{overpic}
\caption{\small\label{fig-horizon} The left figure depicts a black M2 horizon for which the dipole charge vanishes. Since $\tilde F_4$ is regular and closed, the integration surface ${\cal M}_4$ can be shrunk down to zero size which implies that $Q_\text{M5}=0$. In contrast the fact that polarised antibranes have a nontrivial horizon implies that the dipole charge can be non-zero.}
\end{center}
\end{figure}

A crucial step in our argument above was that $\tilde F_4$ was regular at the horizon and that we could freely transform the integral of $\tilde F_4$ to infinity where it is zero, thereby leading to a contradiction. Once the antibranes polarise to spherical M5 branes with induced anti M2-charge both of these arguments break down. First of all we \emph{do} expect a singular $\tilde F_4$ close to an M5-brane to account for the charge. Secondly, since the topology of the polarised state is non-trivial one cannot freely transform the integral of $\tilde F_4$ to infinity. In fact, there will be obstructions whenever the integration surface ${\cal M}_4$ is non-trivial in homology on the horizon (see figure \ref{fig-horizon}). Note that for M5-branes the horizon is not 7-dimensional as for M2-branes, so three of the directions in the integral 
\begin{equation}
\oint_H (C_3-C_0\mathcal{H}_3)\w \tilde F_4~,
\end{equation}
are \emph{parallel} to the brane worldvolume. It is for this reason that we clarify that the integration surface in eq. \eqref{Q5def} must be non-trivial as for example in figure \ref{fig-horizon} in order to give a non-vanishing contribution. The polarised antibranes have much in common with black ring solutions in five dimensions \cite{Emparan:2001wn} whose thermodynamics was studied in \cite{Copsey:2005se}. The black rings had the surprising feature that the dipole charge entered into the first law. This was understood as a consequence of the horizon not being spherical as was previously assumed in the black hole thermodynamics literature. If $Q_\text{M5}$, which we will denote as the \emph{dipole} charge, is non-vanishing we can obviously satisfy eq. \eqref{smarrextr}.

In the set-up we are considering, we expect the M5-branes to source a component of $C_3$ extending along the three-sphere they are wrapping. In fact there is a very natural way of satisfying the Smarr relation by letting again $C_0|_H=\d C_3|_H=0$ and $C_3=f(\psi)\vol_{S^3}$, so that the gauge potential $C_3$ equals the volume from on the brane times a function $f(\psi)$. Then the potential $\Phi_\text{M5}$ is proportional to $f(\psi_H)$, the function $f(\psi)$ evaluated at the polarisation radius. The Smarr relation \eqref{smarrextr} reduces to 
\begin{equation}
2p T_\text{M2} \e^{3A} = \mathcal{E}=\Phi_\text{M5}Q_\text{M5}= \mu_\text{M5} f(\psi_H)~. 
\end{equation}
Where we used that $Q_\text{M5}$ is the charge of a single M5-brane, $\mu_\text{M5}$.
Comparing to the probe result \eqref{Pol-angle} we can rewrite this as
\begin{equation}
f(\psi_H) = \f{3\pi^2}{8 hm}\frac{p}{M} ~.
\end{equation}
We learn that in order to recover the probe result in the $p/M\rightarrow0$ limit of the backreacted solution, the function $f(\psi_H)$ should scale as $\psi_H^2$.

\section{Black branes}\label{secBlack}
\begin{figure}[h!]
\begin{center}
\begin{overpic}[width=0.7\textwidth,tics=10]{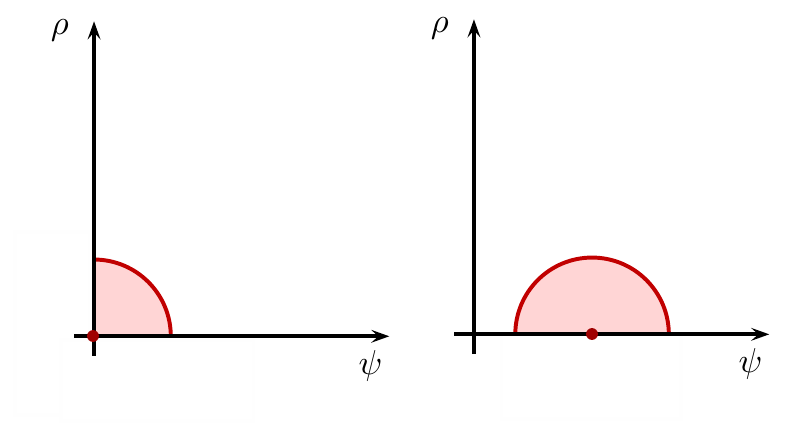}
\end{overpic}
\vspace{-1cm}
\caption{\small\label{fig-horizon-2} The left figure represents a spherical black M2 horizon for which $Q_{\text{M5}}$  vanishes. The right figure depicts a non-extremal M5-brane with induced antibrane charge. For small horizon area we expect the latter to be the dominant phase.}
\end{center}
\end{figure}

After having discussed extremal antibranes, let us take a look at what would be the effect of heating up the system away from extremality. The Smarr relation \eqref{smarr} now also includes non-zero contributions from the area,
\begin{equation}\label{smarrnonext}
{\cal E} = \f76 \f{\kappa {\cal A}}{8\pi G_N} + \Phi_\text{M2}Q_\text{M2} + \Phi_\text{M5}Q_\text{M5}~.
\end{equation}
Remember that 
\be
\f{\kappa {\cal A}}{8\pi G_N} = TS~,
\ee
where $T$ is the temperature and $S$ is the entropy of the black brane. Starting from the extremal state discussed in last section we expect a near-extremal antibrane to have a non-trivial horizon topology. This corresponds to a black M5-brane wrapping a contractible three-cycle on the four-sphere at the tip of the cone. The dipole M5-charge does not vanish (see figure \ref{fig-horizon}) if the topology is non-trivial and $\Phi_\text{M2}$ can be small, or zero as in last section.   
As the horizon area increases we expect an instability towards a collapse to a spherical black brane which cannot support a dipole charge (see figure \ref{fig-horizon}). A regular horizon then demands a cancellation between the form fields
\be
\dd C_3 - C_0 \tilde F_4=0~,
\ee
when restricted to the horizon. This spherical phase, however, does not have a regular extremal limit and so we expect that below some critical area ${\cal A}_\text{crit}$ the dominant phase has non-trivial topology. In figure \ref{fig-horizon-2} we sketch these two phases as horizons in the $\rho-\psi$ plane close to the tip. 

Let us remark that the spherical phase of anti M2-branes (as well as anti D3-branes) was studied in a linear approximation in \cite{Hartnett:2015oda}, where the branes were inserted in a toy model background which captures some of the features of the set-up studied here. There it was observed that the spherical antibranes become singular as the area shrinks to zero size which is consistent with our results

\section{Conclusion}\label{secConclusion}
In this paper we derived the Smarr relation for anti M2-branes (and their polarised state) immersed in the CGLP background \cite{Cvetic:2000db}. We followed a similar procedure as in \cite{Gautason:2013zw,Blaback:2014tfa} where the supergravity equations of motion were combined to find a constraint on the boundary conditions of the solutions at the antibrane location. We showed that these constraints arise when trying to satisfy the Smarr relation \eqref{smarr1}.
We argued that smeared antibranes do not satisfy the Smarr relation without a singular horizon in agreement with \cite{Bena:2010gs,Massai:2011vi,Blaback:2013hqa}. We extended these results to localised extremal anti M2-branes in the CGLP background, and showed that these also cannot be regular while satisfying the Smarr relation. The relation can however be satisfied for an extremal  polarised antibrane, \emph{i.e.} an M5-brane with induced antibrane charge. A crucial feature of the polarised brane is that the dipole M5-charge is nonzero. We do therefore not find a contradiction with the probe results of Klebanov and Pufu \cite{Klebanov:2010qs}. Let us stress that moving away from smeared branes and discussing fully localised branes was crucial to reach this conclusion. Finally, by combining the probe results with ours, we give boundary values for the form fields that could serve as starting points for numerical study of the full supergravity solution. 

We briefly discussed non-extremal antibranes where we expect at least two phases differing in their horizon topology. We argued that an antibrane with trivial horizon area is unstable towards a black ring-like state for a small horizon area. We leave a closer study of the different phases of antibranes in flux backgrounds and their instabilities to future research. The technology used in this paper could also be employed to study black hole microstate backgrounds that make use of antibranes as their method of breaking supersymmetry. The antibrane charge is carried by a non-supersymmetric supertube that polarises and carries dipole charge.  It would certainly be interesting to analyse whether conditions posed by the Smarr relation can be used to evaluate the accuracy of probe calculations for supersymmetric and non-supersymmetric supertubes.

\section*{Acknowledgements}
We are grateful to Thomas Van Riet and Bert Vercnocke for many discussions and motivations during the completion of this work. We also thank Adam Bzowski, Alessandra Gnecchi, Gavin Hartnett, Marjorie Schillo and Matthew Williams and especially Nikolay Bobev for helpful discussions. DCM would like to acknowledge the Becas Chile scholarship programme of the Chilean government. FFG and JD are supported by the National Science Foundation of Belgium (FWO) grant G.0.E52.14N Odysseus and Pegasus. We also acknowledge support from the European Science Foundation Holograv Network.
\appendix
\section{ADM energy for \textit{p}-branes in general backgrounds}\label{appendixadm}
In this appendix we extend the Komar integrals for asymptotically flat black branes of \cite{Townsend:2001rg} to black branes with arbitrary asymptotics.

We will follow the Noether procedure as presented for instance in \cite{Ortin:2004ms}. This method leads to Komar-like integrals and is closely related to the approach of \cite{Abbott:1981ff}, which one can use to calculate the energy of a $p$-brane in $D$ dimensions with an asymptotically flat transverse space \cite{Townsend:2001rg}. The main ideas of \cite{Townsend:2001rg} generalize to non-asymptotically flat $p$-branes in a natural way, by adding a counter-term to the action that takes care of the infinite contribution of the background.

First of all, we consider the solution obtained by dimensionally reducing along the $p$ spatial dimensions of the brane. The reason for doing this is the fact that Komar surface integrals are uniquely defined for $p=0$ (up to a normalisation factor) as opposed to $p\geq 1$, as explained in \cite{Townsend:2001rg}. To this end, we write the metric as a warped product
\begin{equation}
\d s^2_{D}=g_{IJ}(x)\d x^I\d x^J+v(x)g_{mn}(y)\d y^m\d y^n~,
\end{equation}
with $I,J=0,p+1,p+2,\ldots,D-1$ and $m,n=1,2,\ldots,p$. We will assume that the solution is maximally symmetric along the $p$ spatial directions of the brane. In the $D$-dimensional theory, the Einstein-Hilbert term in the action is
\begin{equation}
\frac{1}{16\pi G_N^{(D)}}\int \d^Dx\sqrt{g}R=\frac{1}{16\pi G_N^{(D)}}\int \d^{D-p}x\d^py\sqrt{-g_{D-p}}\sqrt{g_p} ~v^{p/2}R~,
\end{equation}
where $g_{D-p}$ and $g_p$ are the determinants of the $(D-p)$ and $p$-dimensional metrics, respectively. Let us now define
\begin{equation}
\tilde{g}_{IJ}=v^{\frac{p}{D-p-2}}g_{IJ}~.
\end{equation}
For this tilde metric $\tilde{g}_{IJ}$, one has
\begin{equation}
\sqrt{-\tilde{g}_{D-p}}\tilde{R}_{D-p}=\sqrt{-g_{D-p}}v^{p/2}R_{D-p}+(\cdots)~,
\end{equation}
where $(\cdots)$ are terms containing the vector and scalar fields that we get from the metric when dimensionally reducing. With the transformation to $\tilde{g}_{IJ}$ we get the usual Einstein-Hilbert term in the reduced action, with Newton's constants related as usual through
\begin{equation}
G_N^{(D)}=G_N^{(D-p)}\cdot\text{Vol}_p~, \quad\quad \text{Vol}_p=\int \d^py \sqrt{g_p}~.
\end{equation}
Our solution is here just a point-like black hole, for which we can use the Noether approach in order to compute its mass as explained in \cite{Ortin:2004ms}. To do this, we need to find a one-form $\zeta$ such that the combination $(\d\tilde{\eta}+\zeta\wedge\tilde{\eta})$ vanishes asymptotically, and is identically zero for the background metric. Here, $\tilde{\eta}$ is the timelike Killing vector of the tilde metric. To find this $\zeta$, we can can evaluate $\tilde{\eta}$ for the background:
\begin{align}
\tilde{\eta}&=\tilde{g}_{00}\d t=e^{\frac{2(D-2)}{D-p-2}A^B}\d t \\
\d\tilde{\eta}&=\frac{2(D-2)}{D-p-2}\d A^B\wedge \tilde{\eta}~,
\end{align}
where $e^{2A^B}$ is the warp factor of the background solution. This implies that
\begin{equation}
\zeta=-\frac{2(D-2)}{D-p-2}\d A^B.
\end{equation}

We are now in state of calculating the ADM energy. It is given by \cite{Ortin:2004ms,Townsend:2001rg}
\begin{align}
M &= N\oint\tilde{\star}_{D-p}\left(\d\tilde{\eta}+\zeta\wedge\tilde{\eta}\right),\quad \quad N\equiv-\frac{1}{16\pi G_N^{(D-p)}}\frac{D-p-2}{D-p-3}~.
\end{align}
This formula follows from adding a total derivative $\d\tilde{\star}_{D-p}\zeta$ to the original Lagrangian \cite{Ortin:2004ms}, which will serve as a counter-term for the infinite contribution of the background to the energy. We remark that 
\begin{equation}
\tilde{\star}_{D-p}(\d t\wedge \d r)=e^{\frac{p(D-p-4)}{D-p-2}A}\star_{D-p}(\d t\wedge \d r)~.
\end{equation}
Further, we know that $\tilde{\eta}=e^{\frac{2p}{D-p-2}A}\eta$, so our expression for the mass becomes
\begin{equation}
M = N\oint e^{pA}\star_{D-p}\left[\d\eta+\frac{p}{D-p-2}\d (2A)\wedge\eta+\zeta\wedge\eta\right].
\end{equation}
For the next step, we need the following relation:\footnote{We normalise the Hodge operators in different dimensions by demanding $\vol_D=\vol_p\wedge\vol_{D-p}$, where 
\begin{align}
\vol_D&=\sqrt{g_D}\d t\wedge\d r\wedge \d y^1\wedge \cdots\wedge\d y^p\wedge\d x^{p+2}\wedge\cdots\wedge\d x^{D-1}~, \\
\vol_p&=\sqrt{g_p}\d y^1\wedge \cdots\wedge\d y^p~, \\
\vol_{D-p}&=\sqrt{g_{D-p}}\d t\wedge\d r\wedge \d x^{p+2}\wedge\cdots\wedge\d x^{D-1}~.
\end{align}}
\begin{equation}
\star_{D}\d y^1\wedge\cdots\wedge\d y^p\wedge\d t\wedge\d r =e^{-pA}\star_{D-p}\d t\wedge\d r~,
\end{equation}
so that
\begin{equation}
M = N\oint e^{2pA}\star_{D}\d y^1 \wedge\cdots\wedge \d y^p\wedge\left[\d\eta+\frac{p}{D-p-2}\d (2A)\wedge\eta+\zeta\wedge\eta\right]~.
\end{equation}
Finally, recalling our definitions of the one forms associated with the spatial Killing vectors
\begin{equation}
\lambda_i=g_{ii}\d y^i=e^{2A}\d y^i ~,
\end{equation}
we get
\begin{align}
M &= -\frac{1}{16\pi G_N^{(D)}}\oint \star_{D}\left[\d\eta\wedge\lambda_1 \wedge\cdots\wedge \lambda_p+\frac{1}{D-p-3}\d(\eta\wedge\lambda_1 \wedge\cdots\wedge \lambda_p)\right. \nonumber\\
&\hspace{3.7cm} \left.+\xi\wedge\eta\wedge \lambda_1 \wedge\cdots\wedge \lambda_p\right]~,
\end{align}
where we defined
\begin{equation}
\xi\equiv \frac{D-p-2}{D-p-3}~\zeta~.
\end{equation}
For our set-up, we have
\begin{equation}
\xi=\d \log H~.
\end{equation}

\section{Surface gravity and horizon area}\label{appendixArea}
In this appendix we derive the form of the $\kappa\mathcal{A}$ term appearing in \eqref{TSterm} in a general set-up, with that equation being a special case. Let us consider a static metric of the form
\begin{equation}
\d s^2=-e^{2f(r)}\tilde{g}_{00}\d t^2+g_{\mu\nu}\d x^\mu\d x^\nu+e^{-2f(r)}\d r^2+g_{ij}\d x^i\d x^j~,
\end{equation}
where $\mu,\nu=1,\ldots,p$ and $i,j=p+2,\ldots,D-1$. The factor $e^{2f}$ vanishes at the horizon, while the component $\tilde{g}_{00}$ is regular.

For the timelike Killing vector with components $\xi^\mu=\delta^\mu_0$, the surface gravity $\kappa$ is defined as
\begin{equation}
\kappa=\sqrt{\partial_\mu V\partial^\mu V}~, \quad\quad V=\sqrt{-\xi_\mu\xi^\mu}~,
\end{equation}
with both terms evaluated at the horizon. Clearly for the metric at hands
\begin{equation}
V=\sqrt{e^{2f}\tilde{g}_{00}}~,
\end{equation}
so that at the horizon
\begin{equation}
\partial_\mu V\partial^\mu V=g^{rr}(\partial_r V)^2=\frac{(\partial_r e^{2f})^2g^{rr}\tilde{g}_{00}}{4f}=\frac{(\partial_r e^{2f})^2\tilde{g}_{00}}{4}~,
\end{equation}
where we have taken into account the fact that there $e^{2f}\rightarrow0$. Hence
\begin{equation}
\kappa=\frac{1}{2}\sqrt{\tilde{g}_{00}}\partial_r e^{2f}~.
\end{equation}
Next, we have
\begin{equation}
\d \eta=\d(-e^{2f}\tilde{g}_{00}\d t)=-\tilde{g}_{00}(\partial_r e^{2f})\d r\wedge \d t~.
\end{equation}
Therefore
\begin{align}
\star_D~\d \eta\wedge\lambda_1\wedge\cdots\wedge\lambda_p &= -\sqrt{\tilde{g}_{00}}(\partial_r e^{2f})\sqrt{g_p}~\vol_{D-2-p}\\
&= -2\kappa \sqrt{g_p}~\vol_{D-2-p}
\end{align}
at the horizon, so that
\begin{align}
\oint_H\star_D~\d \eta\wedge\lambda_1\wedge\cdots\wedge\lambda_p &= -2\kappa \mathcal{A}_{\text{eff}}~,
\end{align}
with
\begin{equation}
\mathcal{A}_{\text{eff}}=\oint_H\sqrt{g_p}\vol_{D-2-p}~.
\end{equation}
In the main text we avoid writing explicitly the subscript of $\mathcal{A}_{\text{eff}}$.

\section{M2 charge of the CGLP background} \label{CGLPcharge}
The total M2 charge of the background as measured at the UV can be computed by integrating $G_7$ along the base of the cone for large values of the radial coordinate $\rho$. Asymptotically, from \eqref{background-warp} we see that
\begin{equation}
H(\rho)\approx c_0 + 2^{\frac{3}{2}}3^{\frac{3}{4}}m^2\rho^{-\frac{9}{4}}~.
\end{equation}
At the UV, the component $F_7$ of $G_7$ with all its legs on the base of the cone is
\begin{equation}
F_7\approx2^{\frac{1}{2}}3^{\frac{11}{4}}m^2\rho^{-\frac{13}{4}}\star_8\d\rho~.
\end{equation}
As explained in \cite{Klebanov:2010qs}, it is useful to perform the coordinate transformation
\begin{equation}
\rho=\frac{3^{\frac{1}{3}}}{4}r^{\frac{8}{3}},
\end{equation}
in which the metric of the cone becomes $\d s_8^2=\d r^2+r^2\d V^2_{5,2}$. Then we find
\begin{equation}
F_7\approx2^{7}3m^2r^{-7}\star_8\d r~.
\end{equation}
From the form of the metric, it is clear that $\star_8\d r=r^7\text{vol}_{V_{5,2}}$. The volume of the base is calculated in \cite{Bergman:2001qi}, and it turns out to be equal to $3^3\pi^4/2^7$. Therefore the total M2 Maxwell charge of the CGLP background is
\begin{equation}\label{M2charge}
Q_{\text{CGLP}}=\f{\mu_\text{M2}}{(2\pi\ell_p)^6}\int_{V_{5,2}} F_7=\mu_\text{M2}\f{81\pi^4m^2}{(2\pi\ell_p)^6} = \mu_\text{M2}\frac{M^2}{4}~.
\end{equation}


\section{ADM energy for D-branes}\label{appendixDbranes}
In this appendix we present a general derivation of the ADM energy for D$p$-branes immersed in flux backgrounds of type II supergravity with $p+1$-dimensional maximally symmetric spacetime. We assume a background three-form flux $H$ and $(6-p)$-form flux $F_{6-p}$ which are internal and support a smooth asymptotically Ricci flat metric. We also allow for a fluctuating internal $(8-p)$-form $F_{8-p}$. Asymptotically AdS metrics can be treated in a similar way as was done in the main text.

Once the D$p$-brane is introduced into the game, we expect a backreaction onto the metric and the form fields.
The metric splits into the worldvolume metric and a transverse part
\be
\dd s^2 = \e^{2A}\left(-\e^{2f}\dd t^2 + \dd x_p^2 \right) + \dd s_{9-p}^2~,
\ee
where $t$ and $x_p$ span the worldvolume coordinates of the antibrane.

The trace reversed Einstein equation (in Einstein frame) along the brane worldvolume is
\be
R_{\mu\nu} = -\f1{16}\left(2\e^{-\phi}|H|^2 + (7-p)\e^{\f{p-3}{2}\phi} |F_{8-p}|^2 + (5-p) \e^{\f{p-1}{2}\phi} |F_{6-p}|^2\right) g_{\mu\nu}
\ee
The form field equations can be written in terms of the magnetic dual forms
\begin{eqnarray}
\dd F_{p+2} &=& 0~,\\
\dd F_{p+4} - H\w F_{p+2} &=& 0~,\\
\dd H_7 + \eta F_{6-p}\w \sigma(F_{p+2}) &=& 0~,
\end{eqnarray}
where $\eta=(-1)^p$ and the operator $\sigma$ reverses all form indices. The forms in these equations are related to the ones in the Einstein equation by the usual duality rules
\be
H_7=\e^{-\phi}\star_{10} H~,\quad F_{p+2}= \e^{\f{p-3}{2}\phi}\star_{10}\sigma (F_{8-p})~,\quad F_{p+4}=\e^{\f{p-1}{2}\phi}\star_{10}\sigma (F_{6-p})~.
\ee
Using the form equations above we write a set of globally defined gauge potentials:
\begin{eqnarray}
F_{p+2} &=& -\eta~\sigma(\vol_{p+1})\w \dd A_0~,\\
F_{p+4} &=& -\eta~\sigma(\vol_{p+1})\w \left[\dd A_2 + H A_0\right]~,\\
H_7 &=& \eta~\vol_{p+1}\w \left[\dd A_{5-p} - \eta F_{6+p} A_0\right]~.
\end{eqnarray}
The existence of these potentials is not affected by the presence of the anti-D$p$ brane or its polarised states. 
We can now rewrite the right hand side of the Einstein equation as
\begin{eqnarray}
\star_{10}1 \left(2\e^{-\phi}|H|^2 + (7-p)\e^{(p-3)\phi/2} |F_{8-p}|^2 + (5-p) \e^{(p-1)\phi/2} |F_{6-p}|^2\right) \nonumber\\= 
-\vol_{p+1}\w\dd\left(-2 A_{5-p}\w H - (7-p) A_0~ F_{8-p} - (5-p) A_2\w F_{6-p}\right)~.
\end{eqnarray}
Since we want to end up with a ADM energy \emph{density} we redefine the potentials we work with. Finally using the form of the metric the worldvolume Einstein equations can now be written as two PDEs
\begin{eqnarray}
\dd\left(\e^{(p+1)A +f}\star_{9-p}\dd f\right) &=& 0~,\label{genpde1}\\
\dd\left(\e^{(p+1)A + f}\star_{9-p}\dd A - {\cal B}\right) &=& 0~.\label{genpde2}
\end{eqnarray}
where
\be
{\cal B} = -\f1{16}\left[2 A_{5-p}\w H + (7-p) A_0~ F_{8-p} + (5-p) A_2\w F_{6-p}\right]~.
\ee

Evaluating the general ADM energy density formula \eqref{genEnergy} we obtain
\be
{\cal E} = \f{1}{16\pi G_N} \f{1}{7-p}\oint_\infty \e^{(p+1)A + f}\star_{9-p}\left[16\dd A+2(8-p)\dd f\right]~.
\ee
The equations (\ref{genpde1}-\ref{genpde2}) allow us to move the integration surface down to the horizon
\bea
16\pi G_N {\cal E} &=& \f{1}{7-p}\oint_H \left\{\e^{(p+1)A + f}\star_{9-p}\left[16\dd A+2(8-p)\dd f\right] + {16\cal B}\right\} - \f{1}{7-p}\oint_\infty {16\cal B}
\eea
At infinity we expect that $A_{5-p},A_2\to 0$ while the $A_0\to 1$. The last term will therefore give the D$p$ charge of the background, and we can normalise this away in the same way as in the main text by including a counter-term in the action. Here we will simply drop this finite term from the expression. The first term in the integrand gives the surface gravity times the area as explained in appendix \ref{appendixArea}. This leaves us with
\be\label{finex}
{\cal E} = \f{8-p}{7-p}\f{\kappa {\cal A}}{8\pi G_N} + \f{1}{16\pi G_N}\oint_H \f{16\cal B}{7-p}~.
\ee
At this stage we define local gauge potentials at the horizon
\be
\dd B_2 = H~,\quad \dd B_{5-p} = \eta F_{6-p}~.
\ee
These can be used to rewrite ${\cal B}$ at the horizon
\be
16{\cal B} =  -2 (A_{5-p}+A_0B_{5-p})\w H - (7-p) A_0~ \tilde F_{8-p} - (5-p) (A_2+A_0B_2)\w F_{6-p}~,
\ee
where
\be
\tilde F_{8-p} = F_{8-p}-\f{5-p}{7-p}B_2\w F_{6-p}- \f{2}{7-p}B_{5-p}\w H~,
\ee
and is closed. The Einstein equations imply that on a regular horizon\footnote{Here we make use of the fact that the D-branes in question are at finite temperature, which regularises their horizon. All extremal D-branes have singular horizon except for the D3-brane.}
\be
H_7~,\quad F_{p+4}~,\quad F_{p+2}\to 0
\ee
which implies that the forms
\be
A_0~,\quad A_{5-p}+ A_0B_{5-p}~,\quad A_2+A_0B_2
\ee
are closed when restricted to the horizon. This implies that on the horizon we can write
\be
A_{5-p}+ A_0B_{5-p} = \omega_{5-p}+ \text{exact}~,\quad A_2+A_0B_2= \omega_{2}+ \text{exact}~,
\ee
where $\omega_{5-p}$ and $\omega_2$ are harmonic. Repeating the same arguments as in the main text, \emph{i.e.} identifying the Poincar{\'e} duals of the harmonic forms and defining the chemical potentials as their proportionality factors, we are left with
\begin{equation}
{\cal E} = \f{8-p}{7-p}\f{\kappa {\cal A}}{8\pi G_N}+ \Phi_{\text{D}p}Q_{\text{D}p} + \Phi_{\text{D}(p+2)}Q_{\text{D}(p+2)}+ \Phi_\text{NS5}Q_\text{NS5}~.
\end{equation}
All the terms in this expression are analogous to the ones we discussed in the main text. $Q_{\text{D}p}$ is the Page charge, defined as the integral of $\tilde F_{8-p}$ over the horizon.

\bibliographystyle{utphys}
\begingroup
    \setlength{\bibsep}{6pt}
    \setstretch{1}
    \bibliography{references}
\endgroup

\end{document}